\documentclass[aps,preprintnumbers,amsmath,amssymb,superscriptaddress,floatfix,nofootinbib]{revtex4}

\usepackage{lipsum}
\usepackage{graphicx}
\usepackage{epsfig}
\usepackage{epstopdf}
\usepackage{hyperref}
\usepackage{amsmath}
\usepackage{amsfonts}
\usepackage{amssymb}

\usepackage{subfig}
\usepackage{slashed}
\usepackage{soul}
\usepackage[normalem]{ulem}
\usepackage{float}
\usepackage{subfig}
\usepackage{epsfig}
\usepackage{pstricks}
\usepackage{color}

\begin{document}


\title{On the binding of the $BD\bar{D}$ and
$BDD$ systems}

\author{J.~M.~Dias}
\email{jorgivan.morais@ific.uv.es}
\affiliation{Departamento de
F\'{\i}sica Te\'orica and IFIC, Centro Mixto Universidad de
Valencia-CSIC Institutos de Investigaci\'on de Paterna, Aptdo.
22085, 46071 Valencia, Spain}
\affiliation{Instituto de F\'{\i}sica, Universidade de S\~{a}o Paulo,
C.P. 66318, 05389-970 S\~{a}o Paulo, SP, Brazil}

\affiliation{Institute of Modern Physics, Chinese Academy of Sciences, Lanzhou 730000, China}

\author{V.~R.~Debastiani}
\email{vinicius.rodrigues@ific.uv.es}
\affiliation{Departamento de
F\'{\i}sica Te\'orica and IFIC, Centro Mixto Universidad de
Valencia-CSIC Institutos de Investigaci\'on de Paterna, Aptdo.
22085, 46071 Valencia, Spain}

\affiliation{Institute of Modern Physics, Chinese Academy of Sciences, Lanzhou 730000, China}

\author{L.~Roca}
\email{luisroca@um.es}
\affiliation{Departamento de F\'isica, Universidad de Murcia, E-30100 Murcia, Spain}

\author{S.~Sakai}
\email{shuntaro.sakai@ific.uv.es}
\affiliation{Departamento de
F\'{\i}sica Te\'orica and IFIC, Centro Mixto Universidad de
Valencia-CSIC Institutos de Investigaci\'on de Paterna, Aptdo.
22085, 46071 Valencia, Spain}

\author{E.~Oset}
\email{oset@ific.uv.es}
\affiliation{Departamento de
F\'{\i}sica Te\'orica and IFIC, Centro Mixto Universidad de
Valencia-CSIC Institutos de Investigaci\'on de Paterna, Aptdo.
22085, 46071 Valencia, Spain}

\affiliation{Institute of Modern Physics, Chinese Academy of Sciences, Lanzhou 730000, China}

\date{\today}

\begin{abstract}

We study theoretically the $BD\bar{D}$ and $BDD$  systems
to see if they allow for possible bound or resonant states.
 The three-body interaction is evaluated implementing the Fixed Center Approximation to the Faddeev equations which considers the interaction
of a  $D$ or $\bar{D}$ particle with the components of a $BD$ cluster, previously proved to form a bound state. We find an $I(J^P)=1/2(0^-)$ bound state for the $BD\bar{D}$ system at an energy around $8925-8985$ MeV within uncertainties, which would correspond to a bottom--hidden-charm meson.
In contrast, the $BDD$ system, which would be bottom--double-charm and hence manifestly exotic, we have found hints of a bound state in the energy region $8935-8985$~MeV, but the results are not stable under the uncertainties of the model, and we cannot assure, neither rule out, the possibility of a $BDD$ three-body state.
\end{abstract}

\maketitle

\section{Introduction}

The traditional field of few body, which has basically concentrated on few nucleon systems \cite{Alt,Fonseca,Epelbaum} or nucleons and hyperons \cite{Hiyama} is gradually giving rise to less conventional systems. Systems with two mesons and one baryon were studied in \cite{19} with the surprising result that the $1/2^+$ low lying excited baryons could be reproduced with this picture. Similar conclusions were found in \cite{22,23}. Systems of three mesons were also studied and many known resonances could be described within this picture \cite{24,25,26}.

The jump to the charm sector was done with the study of the $DNN$ system in Ref. \cite{30} and $NDK$, $\bar{K}DN$, $ND\bar{D}$ molecules were also studied in Ref. \cite{31}. The charm sector with three mesons was initiated with the description of the $Y(4260)$ as a resonant state of $J/\psi K\bar{K}$ \cite{MartinezTorres}. A more complete list of works along those lines can be found in Ref. \cite{DKK}. In this latter work the $DKK$ and $DK\bar{K}$ systems were studied and the Fixed Center Approximation to the Faddeev equations (FCA) was used. The FCA assumes that there is a cluster of two particles, in this case the $DK$, which forms the $D_{s0}^*(2317)$ \cite{34,36,37,38,42}, and the third particle rescatters multiply with the two particles of the cluster. Direct comparison of the results for the $NK\bar{K}$ system with the FCA \cite{Xie}, with a variational method \cite{Jido}, or full Faddeev equations \cite{alberKan}, confirms the accuracy of the FCA to deal with these problems when we have one couple that clearly binds, like in this case where the $\bar{K}N$ gives rise to the $\Lambda(1405)$. The application of the FCA to obtain the $K^-d$ scattering length \cite{Kamalov} also leads to results comparable to those obtained using a different field theoretical approach \cite{Rusetsky}.
In the present case the $DKK$, $DK\bar{K}$ systems are substituted by $BDD$ and $BD\bar{D}$ and there are clear analogies also in the results.

The $DK$ system is bound, and the analogous $BD$ system was also found to be bound in the study of Ref. \cite{SakaiRoca}, where a state of $I=0$ and $J^P=0^+$ was found around $7100$ MeV, with binding energy between $20-50$ MeV, which would be the analogous of the $D_{s0}^*(2317)$ as a bound $DK$ system. The second $D$ or $\bar{D}$ can then scatter with the $BD$ components of the cluster and lead eventually to more binding, giving rise to a three-body molecule. In the first case the $DB$ interaction will be attractive but the $DD$ is mostly repulsive and it is unclear what will prevail. In the second case the $\bar{D}D$ will be attractive but the $\bar{D}B$ is attractive in $I=0$ and repulsive in $I=1$, and again it is uncertain what will happen. This is analogous to the $DKK$ and $DK\bar{K}$ systems where we had a similar behaviour, the $K$ playing the role of the $D$ and the $D$ playing the role of $B$. Our final result will reveal that the $BD\bar{D}$ system clearly binds while the case for the $BDD$ is uncertain.

\section{Formalism}

The bulk of the formalism to implement the FCA to evaluate the three-body interaction in the $BD\bar{D}$ and $BDD$  systems is analogous to the $DK\bar{K}$ and $DKK$ case of Ref.~\cite{DKK}. Therefore in the present section we just show the differences and modifications for the present case and refer to Ref.~\cite{DKK} for further details on the formalism.

Let us address first the $BD\bar{D}$ interaction. We need to consider the isospin doublets $(B^+,B^0)$, $(\bar{B}^0,-B^-)$, $(D^+,-D^0)$, $(\bar{D}^0,D^-)$ and the $|BD,~I=0\rangle$ state  given by
\begin{equation}\label{BDI0}
  |BD,~I=0\rangle = -\frac{1}{\sqrt{2}}(B^+D^0 +B^0D^+)
\end{equation}
We then have to consider the interaction with
a $\bar{D}$ to account for the $BD\bar{D}$ dynamics. Considering
the possible intermediate steps in the multiple scattering, we need the following channels contributing to the three-body interaction:
\begin{center}
\begin{tabular}{c c c}
  $1)D^-[B^+D^0]\quad$ & $2)D^-[B^0D^+]\quad$ & $3)\bar{D}^0[B^0D^0]$ \vspace{5pt} \\
  $4)[B^+D^0]D^-\quad$ & $5)[B^0D^+]D^-\quad$ & $6)[B^0D^0]\bar{D}^0$ \\
\end{tabular}
\end{center}

\begin{figure}[H]\centering
\includegraphics[width=\textwidth]{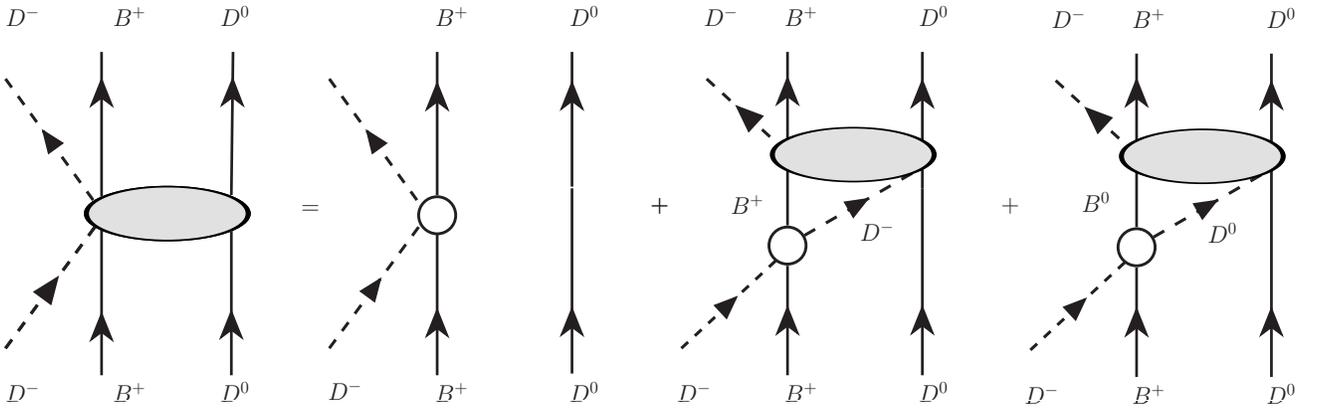}
\caption{Multiple scattering diagrams that go in the construction of the partition function $T_{11}$.}\label{diagT11}
\end{figure}

The difference between the configurations $1)$, $2)$, $3)$ and $4)$, $5)$, $6)$, respectively, is that in the former ones the $\bar{D}$ outside the cluster interacts with the $B$ inside the cluster while in the later ones it interacts with the $D$.
Then we define the partition functions $T_{ij}$ which sum all possible diagrams that begin with configuration $i)$ and finish with configuration $j)$, following an analogous scheme to the one proposed in Ref. \cite{Sekihara}. We show in Fig. \ref{diagT11} the diagrams that contribute to $T_{11}$.

We then have
\begin{equation}\label{T11}
  T_{11}^{\rm FCA} (s) = t_1 + t_1 \,G_0 \,T_{41}^{\rm FCA} + t_2 \,G_0 \,T_{61}^{\rm FCA},
\end{equation}
where $s$ is the total three-body center-of-mas energy; $t_1$, $t_2$ are defined later in Eq. \eqref{tiBDDbar} and  $G_0$ is the $\bar{D}$ propagator modulated by the $BD$ wave function (see details in Eqs. $(2)-(5)$ of Ref. \cite{DKK}, substituting $D \rightarrow B$, $K \rightarrow D$ and $D_{s0}^*(2317)$ by the $BD(7100)$ molecule). We can write the equivalent equations for the other $T_{ij}$ partitions and obtain the set of algebraic equations.
\begin{equation}\label{Tij}
  T_{ij}^{\rm FCA} (s) = V_{ij}^{\rm FCA}(s) +\sum\limits_{l=1}^{6}
 \tilde{V}^{\rm FCA}_{il}(s)\,G_0(s)\,T^{\rm FCA}_{lj}(s)\, ,
\end{equation}
which solution is:
\begin{equation}\label{Tij-1}
T_{ij}^{\rm FCA}(s)=\sum\limits_{l=1}^{6}\Big[\,1-
\tilde{V}^{\rm FCA}(s)\,G_0(s)\,\Big]^{-1}_{il}\,V_{lj}^{\rm FCA}(s)\, ,
\end{equation}
with
\begin{equation}\label{VBDDbar}
  V^{\rm FCA} =
  \left (
  \begin{array}{@{\,}cccccc@{\,}}
    t_{1} & 0 & t_{2} & 0 & 0 & 0 \\
    0 & t_{3} & 0 & 0 & 0 & 0 \\
    t_{2} & 0 & t_{4} & 0 & 0 & 0 \\
    0 & 0 & 0 & t_{5} & 0 & 0 \\
    0 & 0 & 0 & 0 & t_{6} & t_{7} \\
    0 & 0 & 0 & 0 & t_{7} & t_{8} \\
  \end{array}
  \right ) ,
  \qquad
  \tilde{V}^{\rm FCA} =
  \left (
  \begin{array}{@{\,}cccccc@{\,}}
    0 & 0 & 0 & t_{1} & 0 & t_{2} \\
    0 & 0 & 0 & 0 & t_{3} & 0 \\
    0 & 0 & 0 & t_{2} & 0 & t_{4} \\
    t_{5} & 0 & 0 & 0 & 0 & 0 \\
    0 & t_{6} & t_{7} & 0 & 0 & 0 \\
    0 & t_{7} & t_{8} & 0 & 0 & 0 \\
  \end{array}
  \right ) ,
\end{equation}
and the amplitudes $t_i$ are given by
\begin{align}\label{tiBDDbar}
  \begin{tabular}{l l}
    $t_{1}=t_{B^+D^-\to B^+D^-}\,, \quad$ & $t_{5}=t_{D^0D^-\to D^0D^-}\,,$ \\
    $t_{2}=t_{B^+D^-\to B^0\bar{D}^0}\,, \quad$ & $t_{6}=t_{D^+D^-\to D^+D^-}\,,$\\
    $t_{3}=t_{B^0D^-\to B^0D^-}\,, \quad$ & $t_7=t_{D^+D^-\to D^0\bar{D}^0}\,,$  \\
    $t_{4}=t_{B^0\bar{D}^0\to B^0\bar{D}^0}\,, \quad$ & $t_8=t_{D^0\bar{D}^0\to D^0\bar{D}^0}\, .$\\
  \end{tabular}
\end{align}
The amplitudes in Eq.~(\ref{tiBDDbar}) are the $B\bar{D}$ and $D\bar{D}$ unitarized scattering amplitudes, taken
from Refs.~\cite{SakaiRoca} and \cite{37,Gamermann:2008jh} respectively. Note that, as explained in Ref.~\cite{37}, there is, with respect to \cite{SakaiRoca,37,Gamermann:2008jh}, an extra normalization factor $M_{BD}/M_i$ with $M_{BD}$ the mass of the bound state found in the $BD$ system \cite{SakaiRoca} and $M_i$ the mass of the particle of the cluster involved.
This is introduced for convenience to use the Mandl-Shaw \cite{MandlShaw:2010} normalization for external $\bar{D} (D)$ and $[BD]$ states.
For the evaluation of $D\bar{D}$ interaction in Ref.~\cite{Gamermann:2008jh}, from where the $X(3700)$ resonance was dynamically obtained,
other meson-meson channels were considered,  like $\pi\pi$, $\eta\eta$, $K\bar K$, $D_s \bar D_s$ and $\eta \eta_c$, but which turn out to have much less influence than the $D\bar{D}$ channel. Hence, we can neglect in the present work all the channels except the  $D\bar{D}$. However, if we do this we cannot get a width for the $X(3700)$ resonance since the $D\bar{D}$ threshold is far above the position of that resonance. Therefore, in order to get also the width of the $X(3700)$ obtained in Ref. \cite{Gamermann:2008jh}, which was 36~MeV, we have included the $\eta\eta$ channel in addition to the $D\bar{D}$ but with a renormalized value of the $\eta\eta\to\ D\bar D$ potential such as to reproduce the 36~MeV width.

On the other hand, for the case $BDD$ case we can proceed in an analogous way and again we get Eq. \eqref{Tij-1} but with
\begin{equation}
\label{VBDD}
  V^{\rm FCA} =
  \left (
  \begin{array}{@{\,}cccccc@{\,}}
    \bar{t}_{1} & 0 & 0 & 0 & 0 & 0 \\
    0 & \bar{t}_{2} & \bar{t}_{3} & 0 & 0 & 0 \\
    0 & \bar{t}_{3} & \bar{t}_{4} & 0 & 0 & 0 \\
    0 & 0 & 0 & \bar{t}_{5} & 0 & \bar{t}_{5} \\
    0 & 0 & 0 & 0 & \bar{t}_{6} & 0 \\
    0 & 0 & 0 & \bar{t}_{5} & 0 & \bar{t}_{5} \\
  \end{array}
  \right ) ,
  \quad
  \tilde{V}^{\rm FCA} =
  \left (
  \begin{array}{@{\,}cccccc@{\,}}
    0 & 0 & 0 & \bar{t}_{1} & 0 & 0 \\
    0 & 0 & 0 & 0 & \bar{t}_{2} & \bar{t}_{3} \\
    0 & 0 & 0 & 0 & \bar{t}_{3} & \bar{t}_{4} \\
    \bar{t}_{5} & 0 & \bar{t}_{5} & 0 & 0 & 0 \\
    0 & \bar{t}_{6} & 0 & 0 & 0 & 0 \\
    \bar{t}_{5} & 0 & \bar{t}_{5} & 0 & 0 & 0 \\
  \end{array}
  \right ) ,
\end{equation}
and
\begin{align}\label{tiBDD}
  \begin{tabular}{l l}
   $\bar{t}_{1}=t_{D^+B^+\to D^+B^+}\,, \quad$ & $\bar{t}_{4}=t_{D^0B^+\to D^0B^+}\,,$  \\
   $\bar{t}_{2}=t_{D^+B^0\to D^+B^0}\,, \quad$ & $\bar{t}_{5}=t_{D^+D^0\to D^+D^0}\,,$  \\
   $\bar{t}_{3}=t_{D^+B^0\to D^0B^+}\,, \quad$ & $\bar{t}_{6}=t_{D^+D^+\to D^+D^+}\,,$ \\
  \end{tabular}
\end{align}
where the $DD$ amplitudes are taken from Ref. \cite{37}. We use isospin symmetry to build the $I=1$ amplitude of $DD$, including a factor $2$ in the interaction of $D^+D^0 \to D^+D^0$ given in Ref. \cite{37}, while the $I=0$ amplitude vanishes. One should be careful to include a factor $1/2$ in the kernel to account for the normalization of identical particles, which later has to be restored multiplying $t_{DD}^{I=1}$ by $2$.

Finally, the  $T_{BD\bar{D}(D)}$ three-body scattering amplitude in isospin $1/2$ in terms of the amplitudes in Eq.~(\ref{Tij-1}) is given by
\begin{align}\label{Ttotal}
\nonumber T_{BD\bar{D}(D)}=&\frac{1}{2}\Big(
T^{\rm FCA}_{11}+T^{\rm FCA}_{12}+T^{\rm FCA}_{14}+T^{\rm FCA}_{15}
+T^{\rm FCA}_{21}+T^{\rm FCA}_{22}+T^{\rm FCA}_{24}+T^{\rm FCA}_{25}\\
&+T^{\rm FCA}_{41}+T^{\rm FCA}_{42}+ T^{\rm FCA}_{44}+T^{\rm FCA}_{45}
+T^{\rm FCA}_{51}+T^{\rm FCA}_{52}+T^{\rm FCA}_{54}+T^{\rm FCA}_{55}
\Big)\,.
\end{align}

\section{Results}

We have two main sources of uncertainty in our model. The first one is the cutoff used to regularize the $BD$ and $B\bar D$ loop functions
needed for the evaluation of the unitarized scattering amplitudes    \cite{SakaiRoca} of Eqs.~(\ref{tiBDDbar}) and (\ref{tiBDD}). This is carried out in Ref.~\cite{SakaiRoca} and in the present work by using a three-momentum cutoff within the range $q_{\rm max}=400-600$~MeV. The second source of uncertainty is the prescription used to evaluate the
center-of-mass energy ($\sqrt{s_{3i}}$) of the projectile, particle $(3)$ ($\bar{D}$ or $D$) and one of the particles in the cluster, $(1)$ or $(2)$ ($B$ or $D$).
These energies are the argument entering the amplitudes in Eqs.~(\ref{tiBDDbar}) and (\ref{tiBDD}).
Two prescriptions were given in Ref. \cite{DKK} (see Eqs.~$(19-22)$ of that reference) and we consider also both of them in the present work.
 The first prescription (I) is standard in works implementing the FCA scheme and was used in Refs.~\cite{40,51}, and the second prescription (II) takes into account the sharing of the binding energy between the three particles and was introduced in Ref.~\cite{DKK}.
We will consider the differences obtained changing the value of $q_{\rm max}$ and implementing both prescriptions for $\sqrt{s_{3i}}$ as an estimation of the uncertainty of our results.

\begin{figure}[H]\centering
\includegraphics[width=0.8\textwidth]{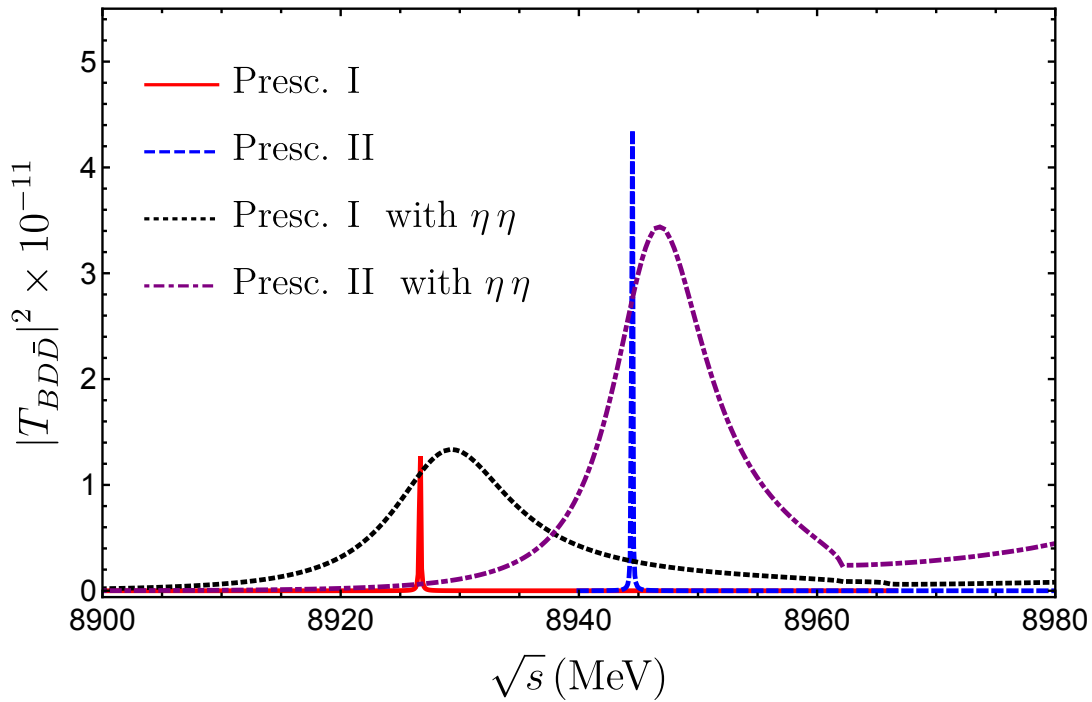}
\caption{$|T_{BD\bar{D}}|^2$ with prescriptions I, II for $\sqrt{s_{\bar{D}B}}$, $\sqrt{s_{\bar{D}D}}$ and $q_{\rm max}=600$~MeV with and without considering width (from $\eta \eta$ channel) for the $X(3700)$ through the $D\bar D$ interaction. The two curves with $\eta \eta$ channel were multiplied by a factor $10^4$ for comparison.}\label{T2BDDbar}
\end{figure}

In Fig. \ref{T2BDDbar} we show the results for $|T_{BD\bar{D}}|^2$ in terms of $\sqrt{s}$, the overall CM energy of the $BD\bar{D}$ three-body system, using both prescriptions for $\sqrt{s_{3i}}$ and $q_{\rm max}=600$ MeV.
We show results with and without considering the width of the $X(3700)$, through the inclusion or not of the $\eta\eta$ channel in the $D\bar D$ interaction, as explained below Eq.~(\ref{tiBDDbar}).
If we do not consider this width, we find, both for prescription I and II,
 a neat and narrow peak  below the threshold corresponding to
 the $BD$ cluster resonant mass ($7093$ MeV) $+$ the $\bar{D}$ mass which equals $8962$~MeV. This peak corresponds, thus, to a three-body bound state. The peaks appear at $8926$ MeV (prescription I) and $8944$ MeV (prescription II).
If we include the $\eta\eta $ channel in the  $D\bar D$ interaction such as to get the right $X(3700)$ width, the position of the states found barely increases by less than 5~Mev but a width for the three-body state is obtained of about 10~MeV.

 The results with $q_{\rm max}=400$ MeV are similar with peaks at the energies $8977$ MeV and $8983$ MeV, respectively, compared with the new threshold of the $BD$ cluster ($7129$ MeV) $+$ $\bar{D}$ mass which equals $8998$~MeV, and once again the $BD\bar{D}$ system binds.
 The different values obtained with the different prescriptions and cutoffs provide a value for the $BD\bar{D}$ bound state in the range $8925-8985$~MeV, with a dispersion in the results that can be considered as the uncertainty of our calculation.

 It is interesting to see the origin of this binding. The cluster $BD$ is bound thanks to the attractive $BD$ interaction in $I=0$. The $B\bar{D}$ is attractive in $I=0$ but repulsive in $I=1$ \cite{SakaiRoca}. Similarly the $D\bar{D}$ is attractive in isospin $I=0$, generating a narrow bound state (the $X(3700)$) around $3720$ MeV \cite{37,Gamermann:2008jh}, and in $I=1$ it is also attractive, but very weakly. In Eq. \eqref{tiBDDbar} one can write the $t_i$ amplitudes in terms of the isospin amplitudes and see that both $I=0$ and $I=1$ $B\bar{D}$ and $D\bar{D}$ amplitudes participate in the process. In order to illustrate the importance of the most attractive components we remove the $I=0$ part of the $B\bar{D}$ amplitudes or the $I=0$ part of the $D\bar{D}$ ones. If we remove the $I=0$ part of the $B\bar{D}$ interaction the peak disappears for both prescriptions and for the two cutoffs. If we only remove the $I=0$ part of the $D\bar{D}$ interaction, then the peak only disappears in prescription II with cutoff $400$ MeV; in all the other cases a peak still remains, although the binding becomes smaller. We conclude that, while the $D\bar{D}$ attraction helps in the building up of the three-body bound state, the main source of binding is the $I=0$ component of the $B\bar{D}$ interaction.

\begin{figure}[h!]
  \subfloat[]{%
    \includegraphics[width=0.48\textwidth]{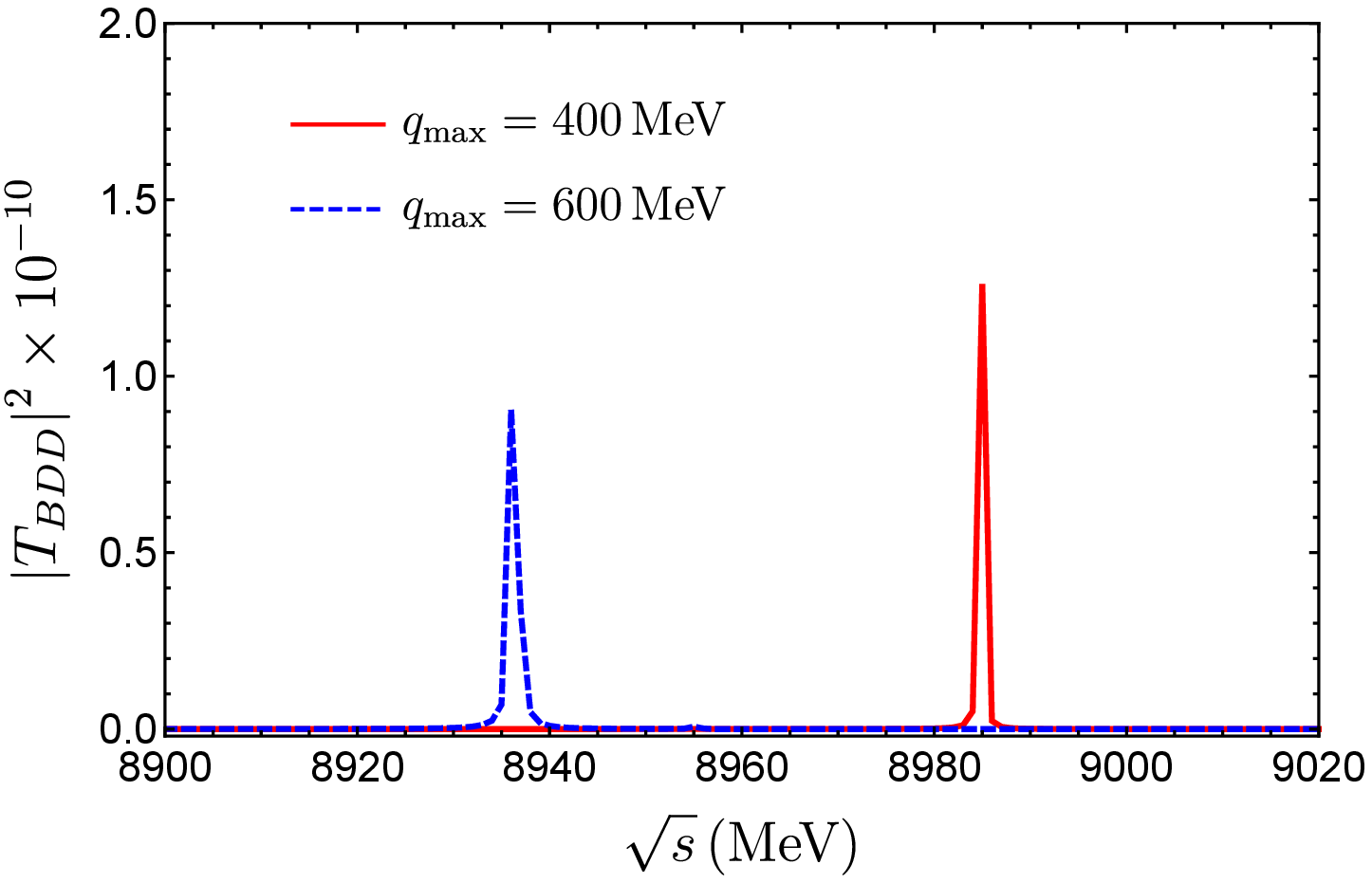} \label{T2BDDpresc1}
  }
  \quad
  \subfloat[]{%
    \includegraphics[width=0.48\textwidth]{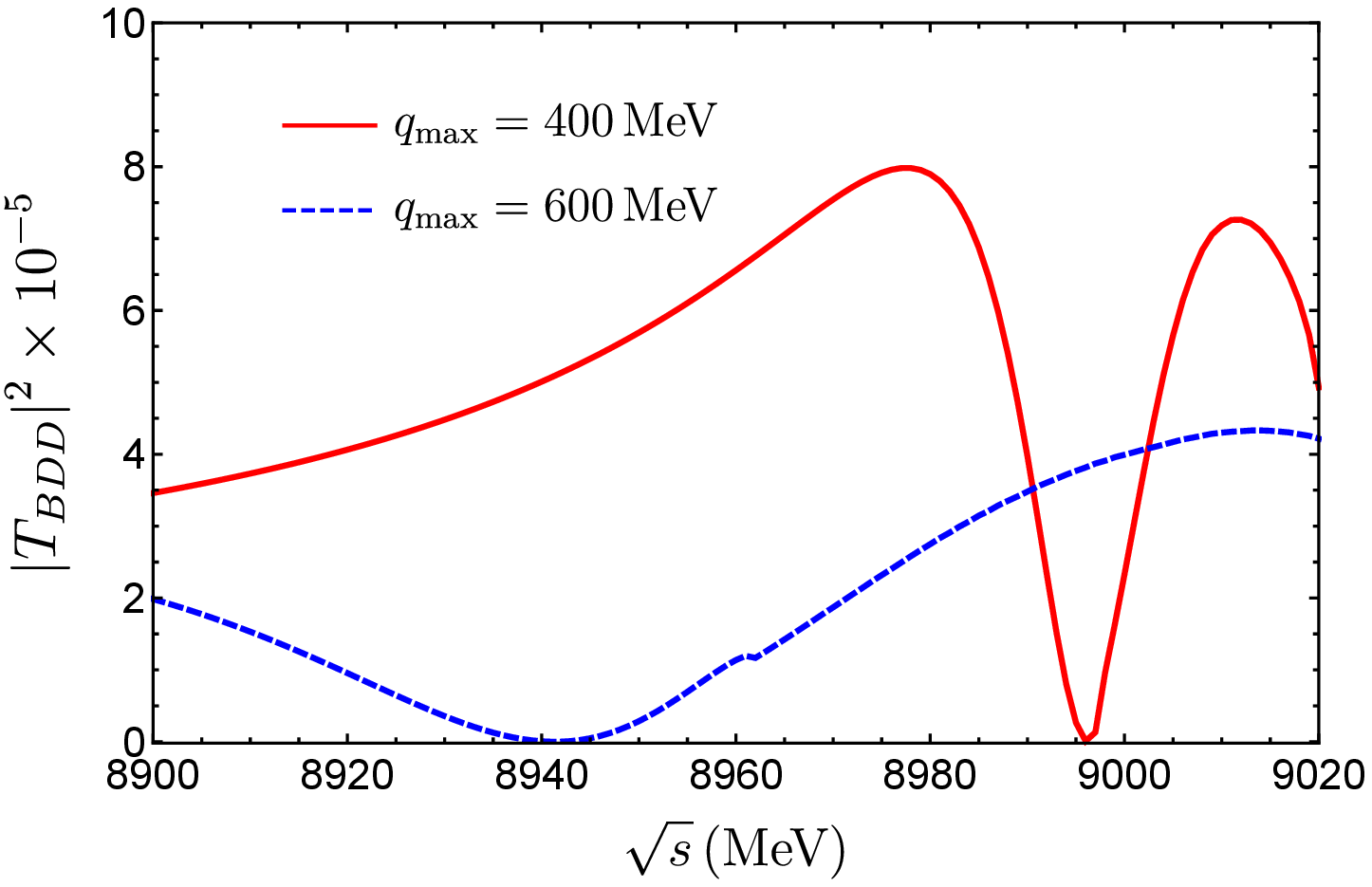} \label{T2BDDpresc2}
  }
  \caption{$|T_{BDD}|^2$ with $q_{\rm max}=400$ MeV and $q_{\rm max}=600$ MeV for the $BD$ cluster, using prescription I (left) and prescription II (right) for $\sqrt{s_{DB}}$, $\sqrt{s_{DD}}$.}
\end{figure}

 As for the $BDD$ system, which would lead to a manifestly exotic meson with two charm quarks and a bottom antiquark, the amplitudes that we obtain using prescription I show a clear narrow bound state with both cutoffs for the $BD$ cluster, as can be seen in Fig.~\ref{T2BDDpresc1}. For $q_{\rm max}=400$ MeV the peak appears at $8985$ MeV, below the corresponding threshold of $BD$ cluster + $D$ ($8998$~MeV), while for $q_{\rm max}=600$~MeV the peak appears at $8936$ MeV, again below its threshold ($8962$~MeV).

 However, when we switch to prescription II the bound state disappears, as can be seen in Fig.~\ref{T2BDDpresc2}. In this case, we get similar structures to the one that was found in the $DKK$ interaction \cite{DKK}.


 In order to further explore the possibility of binding in the $BDD$ system, we have tried another potential to describe the $DD$ interaction, in analogy to the ones of $KK$ from Ref. \cite{DKK}, with an extra factor $1/2$ to account for the absence of the $\phi$ exchange, based on the local hidden gauge approach \cite{Bando,Meissner,Nagahiro}. With this interaction a structure similar to a resonant state is found below threshold for prescription I, but we have noticed that it is too sensitive to changes in the cutoff values. On the other hand, the results using prescription II still show structures similar to the one found in the $DKK$ system \cite{DKK}, which cannot be clearly related to a three-body bound state or resonance.

It is interesting to notice that in analogy to the $DKK$ system, we have a very attractive interaction of the external $D$ with the $B$ of the cluster (the same responsible for the binding of the cluster itself), confronted with the repulsive interaction of $DD$. The three-body binding seems to depend on a very delicate equilibrium between these two interactions, and in the particular case of the $BDD$ system we could not arrive to a decisive conclusion which would be stable under the model uncertainties.

\section*{Conclusions}

We have studied the $BDD$ and $BD\bar{D}$ systems using dynamical models for the $BD$, $B\bar{D}$ and $DD$, $D\bar{D}$ interaction, which have been tested in previous works.
Given the strong binding of the $BD$ pair, we use the Fixed Center Approximation (FCA) to the Faddeev equations to evaluate the three-body interaction by considering the multiple rescattering of the external $D$ or $\bar D$ meson with the components of the $BD$ cluster. This scheme has proved its reliability in many other cases  where two of the particles of the three-body system are strongly clusterized.

We obtained that the $BD\bar{D}$ system is bound and we get an energy of the $BD\bar{D}$ system of about $8925-8985$~MeV considering uncertainties. This result is quite stable under changes in the model that we implement to determine the uncertainty. Our study reveals that the $I=0$ $B\bar{D}$ and $D\bar{D}$ amplitudes, which are attractive, are the main reason for the binding of the $BD\bar{D}$ system, in particular the attraction of the $B\bar{D}$ pair.

As for the $BDD$ system, which would be manifestly exotic, we have found some clues of a bound state in the energy region $8935-8985$~MeV, but the results where not stable under the theoretical uncertainties of the model. Therefore, we cannot assure, neither rule out, the possibility of a three-body state in the $BDD$ interaction.

From our results of the $BD\bar{D}$ study, we expect that a bottom mesonic resonance with quantum numbers $I(J^P)=1/2(0^-)$ and mass around $8925-8985$~MeV could be experimentally found in future studies in hadron facilities.

\section*{Acknowledgments}

J.~M.~Dias would like to thank the Brazilian funding agency FAPESP for the financial support.
V.~R.~Debastiani wishes to acknowledge the support from the
Programa Santiago Grisolia of Generalitat Valenciana (Exp. GRISOLIA/2015/005).
This work is also partly supported by the Spanish Ministerio de Economia
y Competitividad and European FEDER funds under the contract number
FIS2014-57026-REDT, FIS2014-51948-C2-1-P, and FIS2014-51948-C2-2-P, and
the Generalitat Valenciana in the program Prometeo II-2014/068.


\bibliographystyle{plain}

\end{document}